# The missing links in the BGP-based AS connectivity maps

Shi Zhou and Raul J. Mondragon

***Abstract*—A number of recent studies of the Internet topology at the *autonomous systems* level (AS graph) are based on the BGP-based AS connectivity maps (*original* maps). The so-called *extended* maps use additional data sources and contain more complete pictures of the AS graph. In this paper, we compare an original map, an extended map and a synthetic map generated by the Barabási-Albert model. We examine the recently reported rich-club phenomenon, alternative routing paths and attack tolerance. We point out that the majority of the missing links of the original maps are the connecting links *between* rich nodes (nodes with large numbers of links) of the extended maps. We show that the missing links are relevant because links between rich nodes can be crucial for the network structure.**

***Index Terms*—Internet, topology, modeling, networks.**

## I. Introduction

A number of recent studies [1], [2], [5], [6], [7], [8], [9] of the Internet topology at the *autonomous systems* level (AS graph) are based on the BGP-based AS connectivity maps ( *original* maps) [3] [13] [14]. These studies include the discovery of the power laws by Faloutsos *et al* [1] and the error and attack tolerance of the Internet by Barabási *et al* [8]. Chen *et al* [4] produced more complete AS maps (*extended* maps), which are constructed using additional data sources and typically have 20-50% or even more links than the original maps. (As we are considering only connectivity in this paper, we do not differentiate between in-degree links and out-degree links.)

In this paper we compare an original map, an extended map and a synthetic map generated by the commonly used simulator of Internet topology, the Barabási-Albert (BA) model [2]. We examine the recently reported rich-club phenomenon [10], alternative routing paths and attack tolerance. We find out that the extended map shows a strong rich-club phenomenon, in which a small number of nodes with large numbers of links (rich nodes) are significantly better connected to *each other* comparing to the original map and the BA model.

Shi Zhou and Raul J. Mondragon are with the Department of Electronic Engineering, Queen Mary, University of London, Mile End Road, London, E1 4NS, United Kingdom. Email: {shi.zhou, r.j.mondragon}@elec.qmul.ac.uk

This research is supported by the U.K. Engineering and Physical Sciences Research Council (EPSRC) under grant no. GR-R30136-01.

The main contribution of this paper is to point out that the majority of the missing links of the original maps are the connecting links *between* rich nodes of the extended maps. We show that the missing links are relevant, because links between rich nodes can be crucial for the network structure.

These results can be useful when choosing additional data sources to effectively improve the completeness of the AS maps and also useful for those modeling the Internet topology and studying its behavior.

## II. Original And Extended AS Maps

Faloutsos *et al* [1] showed that the Internet topology at the *autonomous systems* level (AS graph) has a power law degree distribution $P(k) \sim k^{-\gamma}$, where degree $k$ is the number of links a node has.

Barabási *et al* [2] showed that a power law degree distribution could arise from two generic mechanisms: 1) *growth*, where networks expand continuously by the addition of new nodes, and 2) *preferential attachment*, where new nodes are attached preferentially to nodes that are already well connected. The Barabási-Albert (BA) model (or so-called scale-free model) has generated great interests in various research areas. Many modifications of the BA model have been introduced ever since. And the BA model has been applied to the research on error and attack tolerance of the Internet [8].

The above studies of the power laws and the tolerance of the Internet are mainly based on the BGP-based AS connectivity maps, the *original* maps, which contain connectivity information of the AS graph and are constructed with the BGP routing tables collected by the Oregon route server (*route-views.oregon-ix.net*) [3] [13] [14]. The Oregon route server connects to several operational routers within the Internet for the purpose of collecting BGP routing tables.

Chen *et al* [5] showed that AS maps constructed solely from Oregon route server data contain only a portion of AS connectivity on the Internet. They constructed the *extended* maps [4] of the AS graph using additional data sources, such as the Internet Routing Registry (IRR) data and the Looking Glass (LG) data. The IRR maintains individual ISP's (Internet Service Provider) routing information in several public repositories to coordinate global routing policy. The Looking Glass sites are maintained by individual ISPs to help troubleshoot Internet-wide routing problems. The original maps typically miss 20-50% or even more of the physical links in the extended maps of the AS graph. The degree distributions of the extended maps are heavily tailed, and they deviate from a strict power law.

## III. Topological Differences

We compare an original map, an extended map and a synthetic map generated by the BA model. The original map and the extended map are measured on May 26th, 2001. As shown in Table I, the extended map and the original map have similar numbers of nodes and maximum degrees, whereas the extended map has 40% more links than the original map. The BA model has the similar size of the extended map, but it has a significantly smaller number of maximum degree. We study the topological differences of the three maps by examining the rich-club phenomenon, alternative routing paths and attack tolerance.

### *A. Rich-club phenomenon*

Power law topologies, such as the AS graph and the BA model, show the property that a small number of nodes have large numbers of links. We call these nodes 'rich nodes'. Recently it was reported [10] that the Internet topology shows a *rich-club* phenomenon, in which rich nodes are very well connected to *each other*. Power law topologies can show significantly different degrees of the rich-club phenomenon. In this section we use the rich-club phenomenon to distinguish the structure differences between the networks.

#### *1) Rich-club Coefficient*

To measure how well nodes are connected to each other, we defined the rich-club coefficient. The maximum possible number of links that $m$ nodes can have is $m(m-1)/2$. The node rank $r$ is the rank of a node on a list sorted in a decreasing order of node degree, and $r$ is normalized by the total number of nodes. The rich-club coefficient $\varphi(r)$ is defined as the ratio of the actual number of links over the maximum possible number of links between nodes with node rank less than $r$.

Fig. 1 is a plot of the rich-club coefficient $\varphi(r)$ against node rank $r$ on a logarithmic scale. The plot shows that rich nodes of the two AS maps are significantly better connected to each other than those of the BA model. And rich nodes of the extended map are better connected to each other than those of the original map.

For example, $\varphi(1\%)=32\%$ of the extend map means that the top 1% rich nodes have 32% of maximum possible number of links, comparing with $\varphi(1\%)=17\%$ of the original map and $\varphi(1\%)=5\%$ of the BA model.

#### *2) Link distribution*

We define $l(r_i, r_j)$ as the number of links connecting nodes with node rank $r_i$ and $r_j$, where node rank are divided into 5% bins and $r_i < r_j$. Fig. 2 is a 3D plot of $l(r_i, r_j)$ against corresponding node rank $r_i$ and $r_j$. In all the three power law topologies, the majority of links are links connecting the top 5% rich nodes $l(r_i < 5\%)$. However, the extended map has a

TABLE I
Network Properties

| Properties | Original map | Extended map | BA model |
|---|---|---|---|
| $N$ | 11174 | 11461 | 11461 |
| $L$ | 23409 | 32730 | 34363 |
| $k_{average}$ | 4.2 | 5.7 | 6 |
| $k_{max}$ | 2389 | 2432 | 329 |

$N$ – total number of nodes. $L$ – total number of links.
$k_{average}$ – average degree. $k_{max}$ – maximum degree.

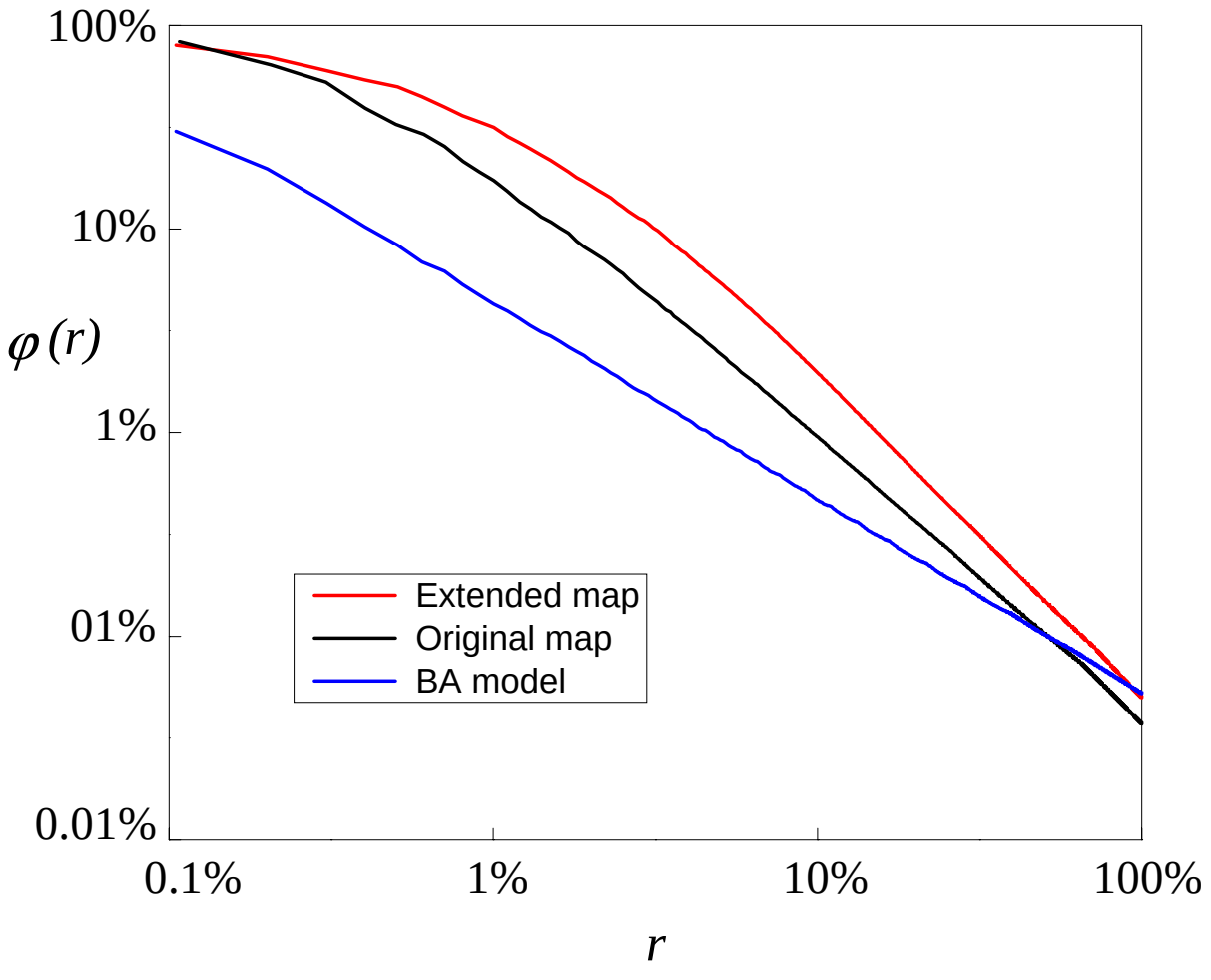

Fig. 1 Rich-club coefficient $\varphi(r)$ against node rank $r$

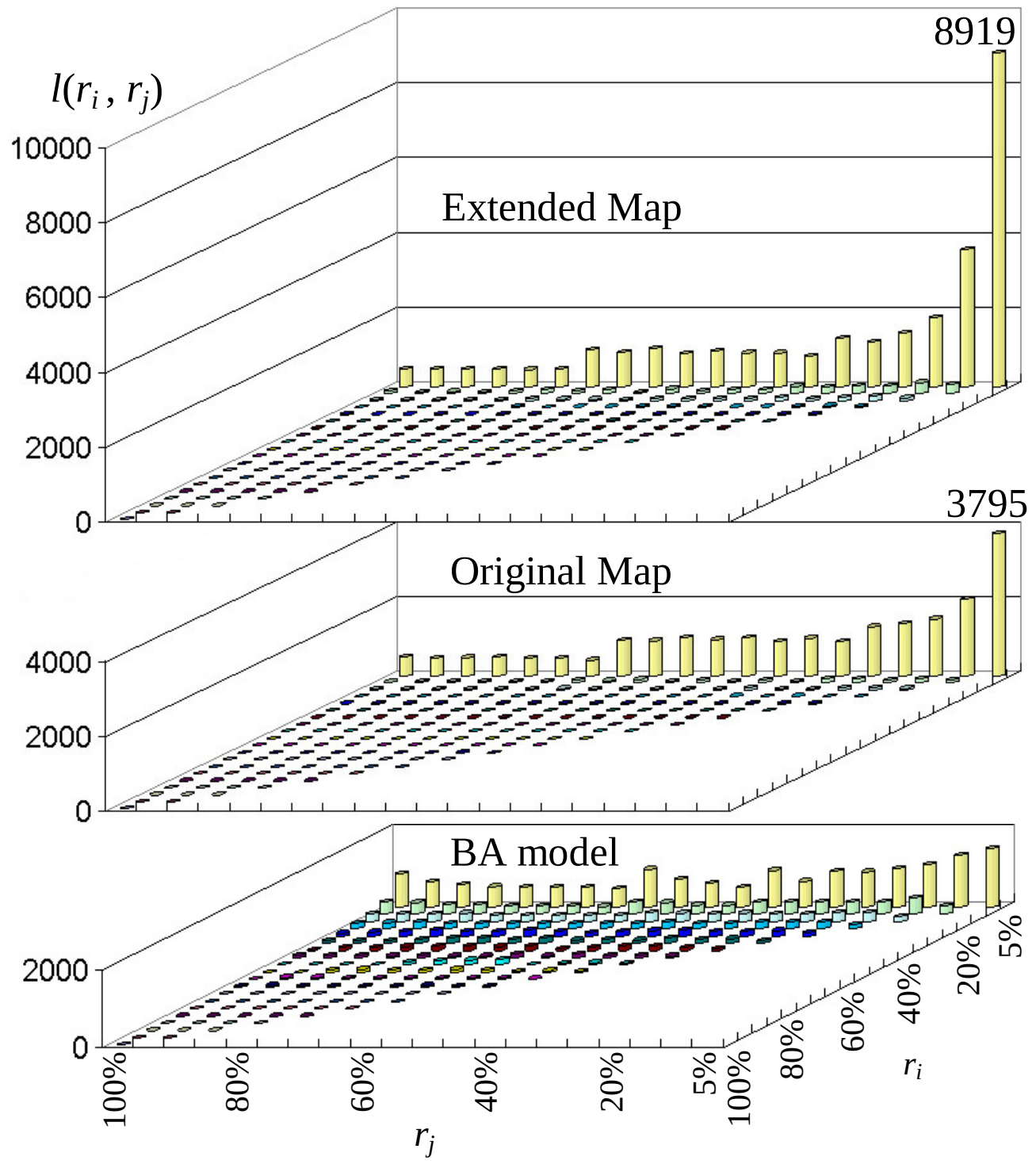

Fig. 2. Number of link $l(r_i, r_j)$ against node rank $r_i$ and $r_j$

TABLE II
ALTERNATIVE ROUTING PATHS

| Properties | Original Map | Extended Map | BA Model |
|---|---|---|---|
| $Kt_{max}$ | 3638 | 7150 | 100 |
| $Kt_{average}$ | 5 | 23 | 0.12 |
| $Kr_{max}$ | 5506 | 8474 | 683 |
| $Kr_{average}$ | 129 | 207 | 1.42 |

$Kt_{max}$ – maximum triangle coefficient. $Kt_{average}$ – average triangle coefficient. $Kr_{max}$ – maximum rectangle coefficient. $Kr_{average}$ – average rectangle coefficient.

significant larger number of connecting links *between* the top 5% rich nodes $l(r_i < 5\%, r_j < 5\%)$. This phenomenon that rich nodes are very well connected with each other is called the rich-club phenomenon. By comparison, the original map does not show the phenomenon as strong as the extended map. Whereas the BA model does not show this phenomenon at all, instead, the top 5% rich nodes are connected to all nodes with similar probabilities regardless of their ranks.

Fig. 2 also shows that the majority of the missing links of the original maps are the connecting links between rich nodes of the extended maps.

## *B. Alternative Routing Paths*

In a topology, a triangle connects a node and two of its neighbor nodes. A rectangle connects a node, two of its neighbor nodes and one of its neighbor's neighbor nodes. The more triangles and rectangles a network has, the more possible alternative routing paths it can have.

The triangle coefficient *Kt* of a node is defined as the number of triangles the node has, and the rank of triangle coefficient *r(Kt)* is defined as the rank of a node on a list sorted in a decreasing order of the triangle coefficient *Kt*.

The rectangle coefficient *Kr* of a node is defined as the number of rectangles the node has, and the rank of rectangle coefficient rank *r(Kr)* is defined as the rank of a node on a list sorted in a decreasing order of the rectangle coefficient *Kr*.

Fig. 3, Fig. 4 and Table II show that the two AS maps have significantly larger numbers of triangles and rectangles than the BA model. And the extended map has more triangles and rectangles than the original map.

## *C. Attack tolerance*

Barabási *et al* [8] shows that it is difficult to divide power law topologies into separate sub networks by removing nodes at random (error), but it is very easy to split them into sub networks by removing specific nodes (attack).

To simulate an attack we first remove the best-connected node, and continue selecting and removing nodes in decreasing order of their degree *k*. In fact, this is equivalent of removing the members of the rich-club. Fig. 5 is a plot of the size of the largest cluster *S*, shown as a fraction of the total system size, against *f* the fraction of the nodes being removed in the attack mode. The figure shows that the extended map and the original map are extremely vulnerable to attack. When the top 5% rich nodes are removed, both AS maps collapse into small pieces. The original map is more fragile than the extended map under attack. Comparing with the two AS maps, the BA model shows a higher degree of attack tolerance.

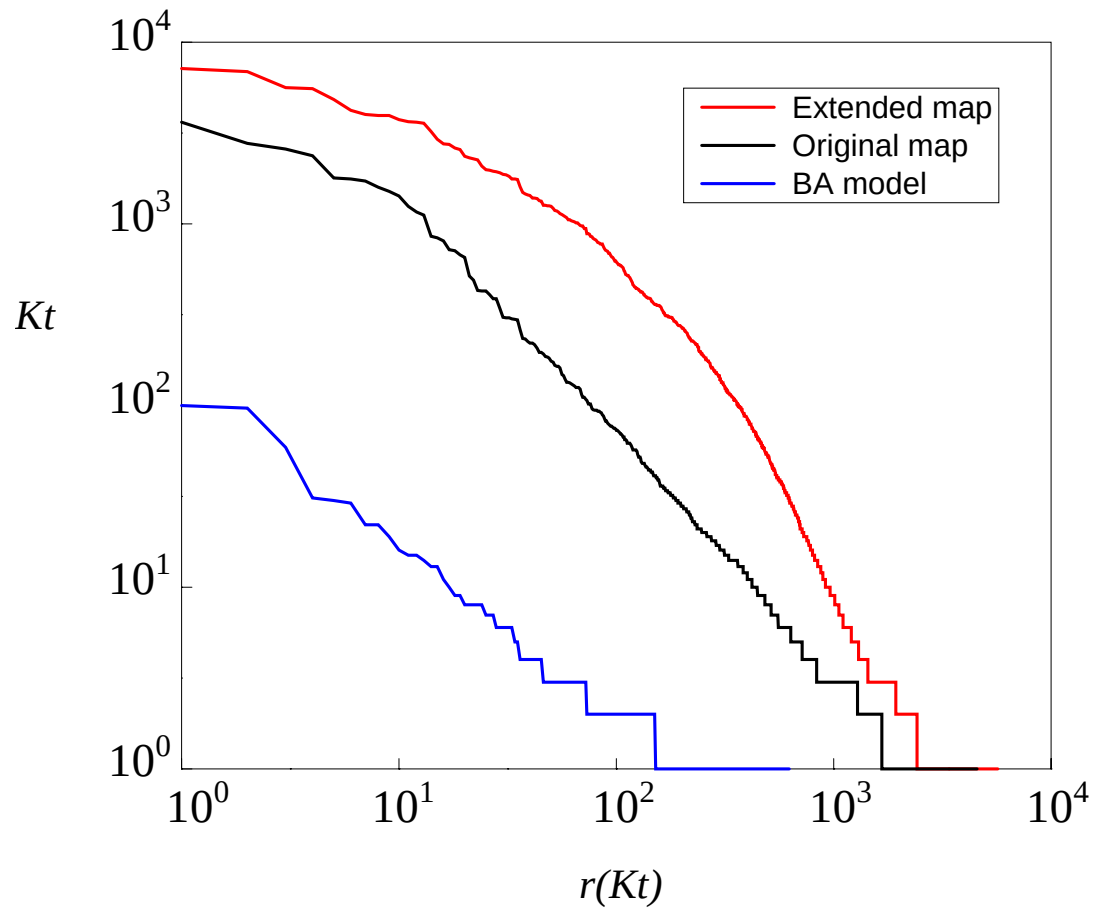

Fig. 3. The triangle coefficient *Kt* against the triangle coefficient rank *r(Kt)*

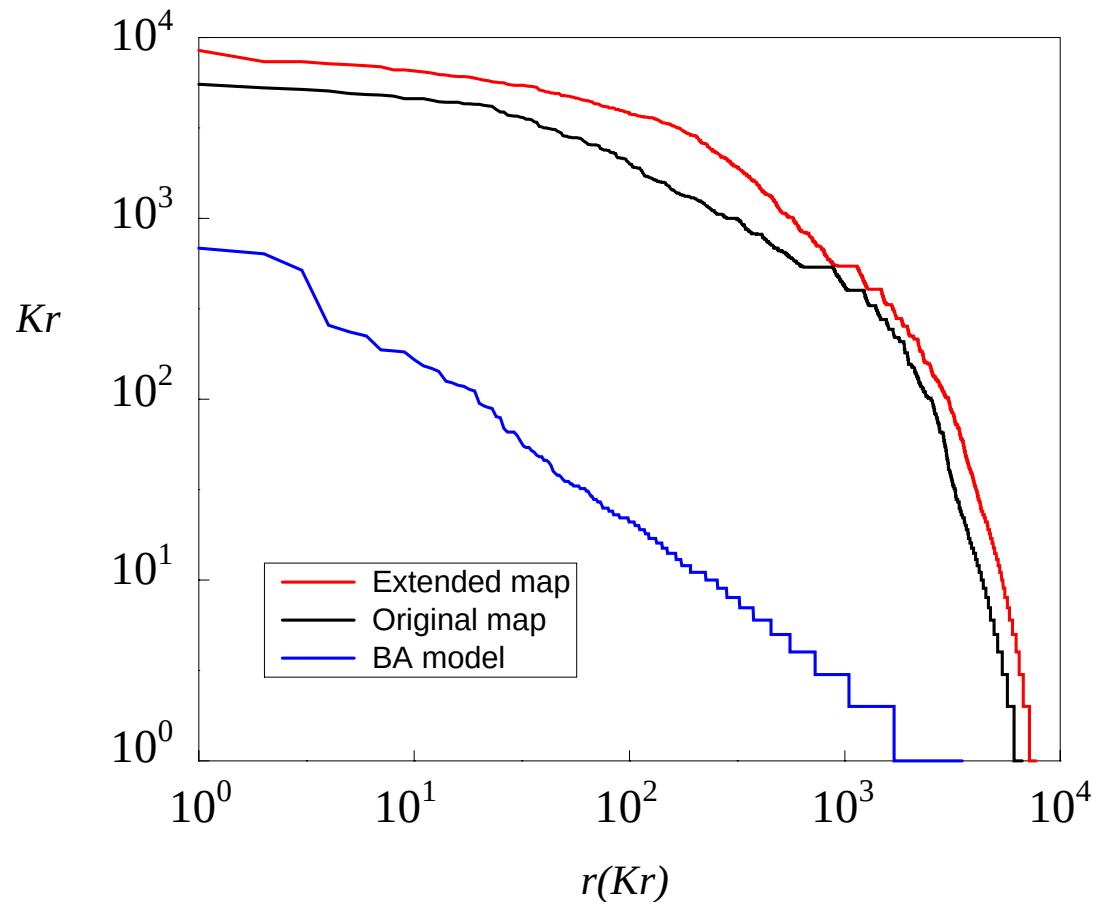

Fig. 4. The rectangle coefficient *Kr* against the rectangle coefficient rank *r(Kr)*

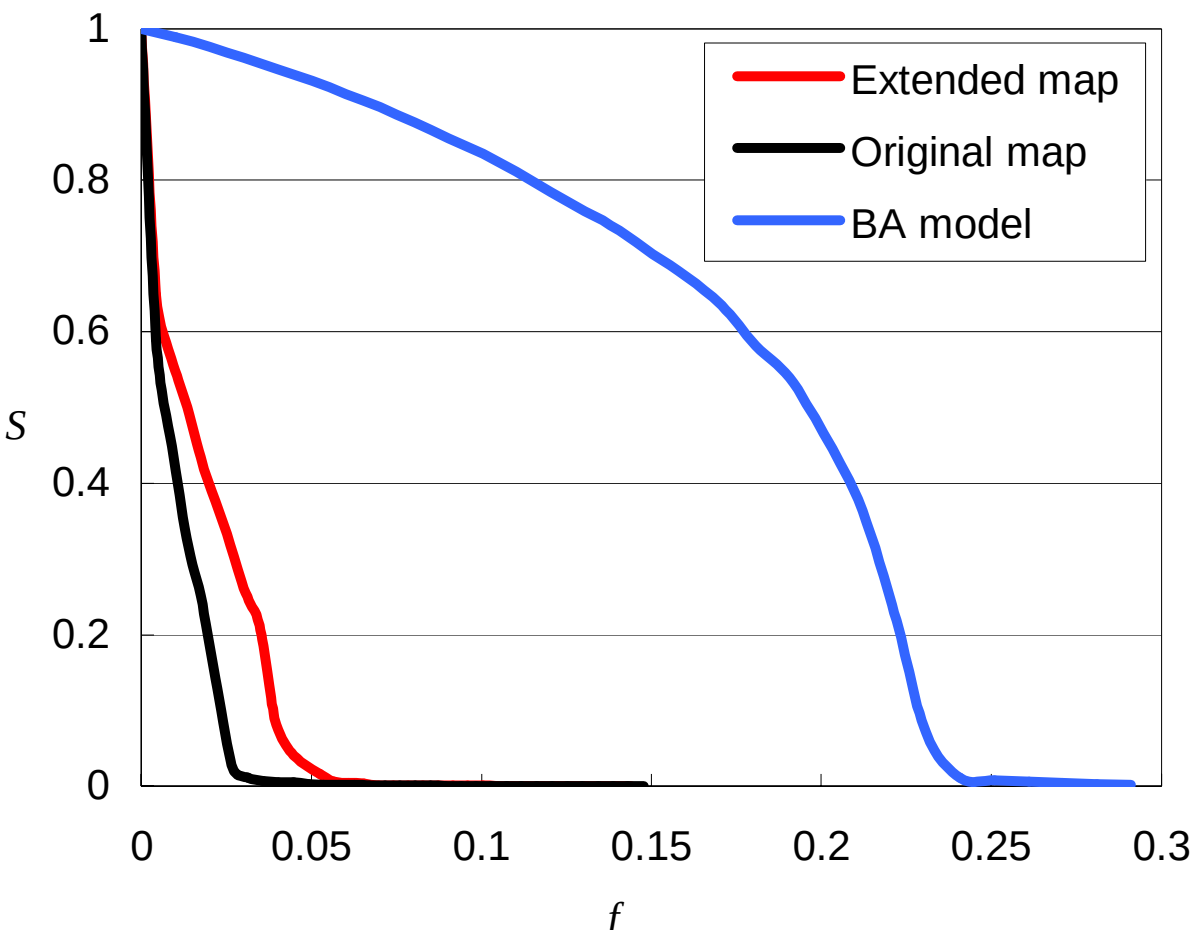

Fig. 5. The size of the largest cluster, *S*, against *f*, the fraction of the nodes being removed in the attack mode.

We have measured the above properties for the AS maps measured on other dates, the main results are still valid.

## IV. Discussion And Conclusion

### A. *The topological role of the missing links*

The extended map shows a rich-club phenomenon stronger than the original map. The majority of the missing links of the original map are the connecting links between rich nodes of the extended map. These links make the rich nodes of the extended map better connected to each other and result in a stronger rich-club phenomenon.

As seen from the comparison between the original map and the extended map, the missing links are relevant, because links between rich nodes can be crucial for the network structure.

### B. *Mapping the Internet*

An explanation of the incompleteness of the original map from a topological perspective is that, because the missing links are mostly the redundant links between rich nodes, they are likely to be the back-up routing paths in the actual Internet. Consequently they are difficult to be fully captured by BGP routing tables, which only provide local views of the global picture.

It is important to keep measuring the Internet, because we still do not know if there are other missing links which are of fundamental importance for the description of the network topology. Our result can be useful when choosing additional data sources to effectively improve the completeness of the AS maps.

### C. *Modeling the Internet*

The BA model has much more links than the original map, but it does not show a rich-club phenomenon at all. Rich nodes of the BA model are connected to all nodes with similar probabilities. This suggests that more links do not necessarily result in the rich-club phenomenon. This result also shows that the topological structure of the BA model is fundamentally different from the AS graph.

The rich-club phenomenon illustrates that networks obeying the same power law degree distribution can have different topological structures. This phenomenon is useful when distinguishing power law topologies and evaluating their generators.

Recently we have introduced the Interactive Growth (IG) model of the Internet topology. This simple and dynamic model is a modification of the BA model and based on the rich-club phenomenon. The IG model compares favorably with other Internet power law topology generators [11].

### D. *Redundancy and robustness*

The extended map has much more links than the original map and these links increase the network redundancy by forming alternative routing paths. Since the rich nodes of the extended map become more dominant in the network, the attack tolerance of the network is only slightly reinforced. This suggests that improved redundancy does not necessarily result in improved robustness.

The BA model has a similar size of the extended map and it has only a few triangles and rectangles. Nevertheless, the BA model shows a significantly higher degree of attack tolerance than the two AS maps. This illustrates that different topological structures have a great impact on network property.

## V. Future Work

Other AS connectivity data sources, such as the RIPE routing information service [12], should be explored to evaluate our result and improve the AS maps. Also our result on attack tolerance shows that a small number of rich nodes are very important for the connectivity of the Internet. It would be interesting to investigate if a change of the local routing policy of one of these nodes has a major impact on the global network.